\documentclass[referee]{aa}
\usepackage{epsf}
\usepackage{graphicx}

\begin{document}

   \title{Angular dimensions of planetary nebulae}


   \author{R. Tylenda\inst{1} \and N. Si\'odmiak\inst{1} 
           \and S. K. G\'orny\inst{1} \and R. L. M. Corradi\inst{2}
           \and H. E. Schwarz\inst{3} }

   \offprints{N. Si\'odmiak, email: alexan@ncac.torun.pl}

   \institute{N. Copernicus Astronomical Center, Department for Astrophysics,
                 Rabia\'nska 8, 87--100 Toru\'n, Poland 
             \and Isaac Newton Group of Telescopes,
                  Apartado de Correos 321, 
                  38700 Sta. Cruz de La Palma, Spain
             \and CTIO/NOAO, Casilla 603, La Serena, Chile}

   \date{Received ; accepted }

   \titlerunning{Angular dimensions of planetary nebulae}
   \authorrunning{Tylenda et al.}

\abstract{
We have  measured angular dimensions of 312 planetary
nebulae from their images obtained in H$\alpha$ (or
H$\alpha$~+~[NII]). We have applied three methods of measurements: direct
measurements at the 10\% level of the peak surface brightness, Gaussian
deconvolution and second-moment deconvolution. The results from the three
methods are compared and analysed. We propose a simple deconvolution
of the 10\% level measurements which significantly improves the reliability
of these measurements for compact and partially resolved nebulae. Gaussian
deconvolution gives consistent but somewhat underestimated diameters compared
to the 10\% measurements. Second-moment deconvolution gives results in poor
agreement with those from the other two methods, especially for poorly
resolved nebulae. From the
results of measurements and using the conclusions of our analysis we derive
the final nebular diameters which should be free from systematic differences
between small (partially resolved) and extended (well resolved) objects 
in our sample.

}

   \maketitle

\section{Introduction}

The angular dimensions belong to the most fundamental
observational data for planetary nebulae (PNe). 
They are required if one wants
to study important PN parameters such as linear dimensions, lifetimes, surface
brightnesses. Several methods for determining distances or estimating
nebular masses rely on observational measurements of the PN diameters. 
Also studies of the
observed evolution of PNe and their central stars often use the angular PN
diameters as one of the crucial observational data.

PN dimensions have been measured since the very begining of
research interest in these objects and
the list of papers with resultant values is
very long. Consequently, for almost all the known PNe we have information
about their angular dimensions, see e.g. the ESO-Strasbourg Catalogue
(Acker et~al. \cite{gpnc}). However the data from many existing catalogues 
and databases should be used with caution. 
They have usually been compiled from numerous different sources based on
different observational techniques and methods. Therefore
the resulting data for different objects are not always comparable.
For many PNe the diameters taken from different observational sources 
can differ by a factor of a few. 

The problem is that different observational techniques and/or different methods
for measuring dimensions give different results for many PNe. In the
case of well resolved and fairly symmetric nebulae 
with a well defined outer rim the problem
is simple and different methods give consistent results. Most of the PNe
have, however, complex structures, soft outer boundaries, and low
surface brightness emission extending to large distances. If one measures
the maximum extension of the nebular emission in an image, as it has often
been done while determining PN dimensions, than the result
depends on how deep the image has been made. A reasonable solution in
order to get results comparable for all objects is to measure nebular
extension at a given level of surface brightness, e.g. at 10\% of the maximum
value.

Another problem arises in the case of small PNe when the nebular diameter is
comparable to the angular resolution of the instrument. The observed image is then a
convolution of an intrinsic nebular image and the instrumental profile.
It has to be deconvolved in order to get a meaningful result. It is usually done 
using the so-called Gaussian deconvolution. The disadvantage of this method
is that one has to adopt a shape of the intrinsic surface brightness distribution
and the result depends on this assumption. The problem of determining
dimensions of partially resolved nebulae has recently been discussed by
van~Hoof (\cite{vh}). He has made a thorough model analysis of different methods
for measuring partially resolved nebulae. 

The question that arises, and which is important for using the measured
dimensions in extensive, systematic studies of PNe, is to what extent the
results from different methods are comparable. For instance, do the
results from direct measurements at the 10\% contour for extended nebulae 
mean the same as the values from a deconvolution method for compact objects? 
If the PNe were close to model nebulae studied by van~Hoof (\cite{vh})
the answer would be: yes, they measure more or less the same. However,
real nebulae are often much different from the model ones and the answer is
not a priori clear. We attempt to answer the above question in this paper.

In recent years, a number of observational surveys
for studying PN structures and dimensions have been done. 
As surveys in radio
wavelengths those of Aaquist \& Kwok (\cite{ak}) and Zijlstra et~al.
(\cite{zpb}) should be mentioned. Over 400 PNe have been observed in these
two surveys, mostly at 5~GHz. PN diameters have been determined either from
a 10\% contour level (for well resolved nebulae) or from a Gaussian
deconvolution (for partially resolved objects). Recent optical surveys can
be found in Schwarz et~al. (\cite{scm}), Manchado et~al. (\cite{mgss}), and
G\'orny et~al. (\cite{gsc}) where about 600 PNe have been observed.
However, the PN dimensions given in these three catalogues usually
refer to the maximum extension of the recorded emission. As
the ratio of the maximum to faintest recorded surface brightness can
vary by a large factor from one image to another these diameters do not
measure the same for all the objects in the catalogues. 

We have measured angular dimensions for 312 PNe. One of the goals is
to provide a homogeneous set of PN diameters, i.e. values which mean the same, 
as much as possible, for all the objects in the sample.  
We have applied direct measurements at 10\% of the peak
surface brightness as well as deconvolution methods. The results have been
used to investigate reliability of the methods and for comparative studies
of the results from different methods. This helped us to obtain a
homogeneous set of the PN diameters for the whole sample without 
important systematic differences between compact and extended nebulae.

\section{Observational material}

The observational material used in this paper is, in principle, the same as in 
Schwarz et~al. (\cite{scm}) and G\'orny et~al. (\cite{gsc}). These are
images of mostly southern PNe taken with the NTT and other telescopes at the
ESO. For our measurements we have used only images in H$\alpha$ (or
H$\alpha$~+~[NII]). This has been done in order to have an uniform set of
data and to avoid possible ionization stratification effects in the results
which could have appeared if we had used, for instance, [OIII] images.

\section{Methods of measurements}

We have applied three methods of measuring PN dimensions. First one is a
direct measurement of the nebular diameter at a level of 10\% of the peak
surface brightness, $\Theta_{10\%}$, from the PN image. 
As the nebulae in majority are not
spherically symmetric we have measured two diameters along perpendicular
axes. One of the axes was directed along the longest
extension of the nebular emission.

Second method is the Gaussian deconvolution. The full width at half-maximum
(FWHM) values, $\Phi$, 
have been obtained from a two-dimensional Gaussian fitting to the observed
PN image. Before fitting the background has been subtracted from the image.
The same applied to field stars allowed us to
derive the FWHM of the resolution profile, $\Phi_{\rm b}$. 
The deconvolved FWHM diameter, $\Phi_{\rm d}$, can then be obtained from
\begin{equation}
\Phi_{\rm d} = \sqrt{\Phi^2 - \Phi_{\rm b}^2}.
\end{equation}
In order to get an estimate of the real nebular diameter, $\Phi_{\rm d}$ has
to be multiplied by a conversion factor which is a function of the
observational resolution, $\Phi_{\rm d}/\Phi_{\rm b}$, and the adopted
intrinsic surface brightness distribution of the nebula. Details of
calculating the conversion factor for a constant surface brightness disc and
constant emissivity shells can be found in van~Hoof (\cite{vh}).

Third method is the so-called second-moment deconvolution. 
This is an analogous method to the Gaussian deconvolution but
the FWHM is determined from the second moment of
the surface brightness distribution. The advantage of this method is that,
unlike Gaussian deconvolution, the conversion factor is independent of the
resolution of the observations. Details of applying second-moment
deconvolution can be found in van~Hoof (\cite{vh}).

It is well known that the first method fails for small objects and therefore
it is usually applied to well resolved nebulae. The deconvolution methods,
on the contrary, are usually applied to small, partially resolved nebulae.
In this paper we have, however, attempted to apply all the three methods
to all the PNe in our sample. 
This has been done in order to be able to compare 
different methods and to study systematic differences 
between results from them.

\section{Results of measurements}

Table 1 presents the results of direct measurements done on our PN images. 
Following suggestions of
van~Hoof (\cite{vh}) we give raw data from the deconvolution methods,
i.e. the deconvolved FWHM, $\Phi_{\rm d}$, and the image resolution,
$\Phi_{\rm b}$. Our final angular diameters of the PNe are
given in Table~2. They are based on the measurements reported in Table~1 as
well as on the discussion of the results from different methods presented in
Sect.~5.

Column (1) in Table~1 gives the PNG
number of the object while its usual name is in column (2). Column (3) gives
the number of a figure showing the PN image in the original papers: "s" and
"g" stand for Schwarz et~al. (\cite{scm}) and G\'orny et~al. (\cite{gsc}),
respectively. The deconvolved FWHM, $\Phi_{\rm d}$, 
and the resolution size, $\Phi_{\rm b}$, from the Gaussian
fit are in columns (4) and (5), respectively. The same but from the
second-moment method can be found in columns (6) and (7). The results of
direct measuremets at 10\% level are given in column (8). Some PNe have 
relatively deep images and the nebular emission extends well beyond the 10\%
contour. In this cases we have also measured the dimensions of the $3\sigma$
emission level above the background. The results are in column (9). Column (10)
gives the percentage of the $3\sigma$ level in the peak surface brightness.
All the measurement results are in arcsec.

For some objects the deconvolution methods do not work well and Table~1 
does not give results in these cases. 
It has usually happened when the nebula was extended with complex
structures and/or had a strong central star or several field stars superimposed 
on the PN image. In
these cases the numerical procedure fitting Gaussian or calculating moments
either has failed or has given spurious results, or has just measured 
a star (central or one of the field stars). Note that the Gaussian fit
fails more often that the second-moment method, particularly for larger
nebulae.

Not all the objects published in Schwarz et~al. (\cite{scm}) 
and G\'orny et~al. (\cite{gsc}) appear in Table~1. Some PNe were
too faint in the images or larger than the image frame so meaningful
measurements have not been possible. On the other hand, images for a few
objects (those with no entry in column (3) in Table~1) have not been published.
For most of these objects their images (although obtained with different equipement) 
can be found in Manchado et~al. (\cite{mgss}).

For a few objects we had two images from independent observations. We give
results from both images separately in two consecutive rows in Table~1 in
these cases.

\section{Comparison of the methods}

In this section we compare the results from the three methods used to
measure the PN dimensions. The aim is to study
systematic effects between different methods. This will allow us to work out a
procedure for deriving a consistent set of PN dimensions for
all the objects, i.e. those well resolved as well as partially resolved, in our
sample.

\subsection{Deconvolution methods: Gaussian versus second-moment}

As discussed above, both, Gaussian fitting and second-moment calculations 
have been applied to
measure the FWHM of the resolution profile, $\Phi_{\rm b}$, on our images.
The results are given in columns (5) and (7) of Table~1.
Both methods should give the same result if the instrumental profile is
Gaussian. As can be seen from Table~1 this is essentially the case.
The mean value and the standard deviation of the ratio
$\Phi_\mathrm{b}(\mathrm{Gauss})/\Phi_\mathrm{b}$(s-m) calculated from
the data in Table~1 is 0.965$\pm$0.094. Thus the basic assumption of
the deconvolution methods, namely Gaussian resolution profile, is
satisfied in our observational material.

\begin{figure}
  \resizebox{\hsize}{!}{\includegraphics{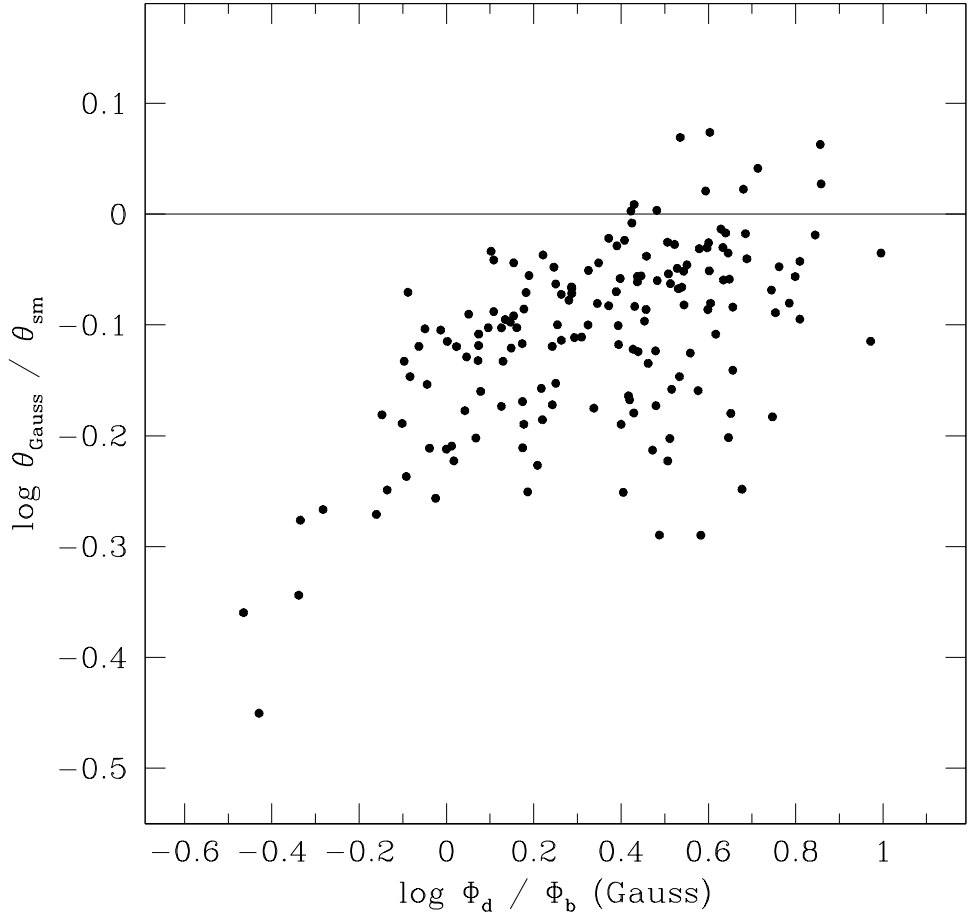}}
  \caption{Comparison of the results from the Gaussian deconvolution and 
           the second-moment deconvolution. The ratio of the angular
           diameters, $\Theta$, derived from the two methods is plotted versus 
           the ratio of the deconvolved FWHM, $\Phi_{\rm d}$, to
           the resolution size, $\Phi_{\rm b}$, obtained from the Gaussian
           fit. A constant emissivity sphere model has been adopted 
           while deriving $\Theta$ in both methods.}
  \label{g_sm}
\end{figure}

The final results of the Gaussian and second-moment
deconvolution methods are compared in Fig.~\ref{g_sm}. 
The figure plots the ratio of the angular diameters, $\Theta$, derived from
the two methods versus the ratio of the deconvolved FWHM, $\Phi_{\rm d}$, to
the resolution size, $\Phi_{\rm b}$, (i.e. parameter $\beta$ in van~Hoof
\cite{vh}) obtained from the Gaussian fit. 
The values plotted in Fig.~\ref{g_sm} are geometric means of the values 
obtained from the two-dimensional measurements reported in Table~1.
While deriving $\Theta$ a surface brightness distribution from 
a constant emissivity sphere model (shell~0.0 in van~Hoof's notation) has been
adopted for calculating the conversion factors. Note,
however, that the diagram shown in Fig.~\ref{g_sm} depends only weakly 
on the adopted surface brightness distribution.

As can be seen from Fig.~\ref{g_sm}, the two deconvolution methods 
do not give the same result
for the bulk of objects. The second-moment method tends to give larger
diameters than the Gaussian deconvolution. In the case of compact nebulae, 
i.e. when the Gaussian deconvolution gives
$\log \Phi_{\rm d}/\Phi_{\rm b}\,\la\,0.0$,
the difference becomes particularly important and increases with
decreasing PN size (as measured from the Gaussian deconvolution).
Note that the vertical scatter of the points in Fig.~\ref{g_sm},
tends to become larger for more extended nebulae.
On average $\log \Theta_{\rm Gauss}/\Theta_{\rm s-m}$ for objects with 
$\log \Phi_{\rm d}/\Phi_{\rm b}\,\ge\,0.0$ is
$-0.096\pm0.071$ (mean and standard deviation). 
When adopting a shell-like model this value would go down and 
for an infinitely thin shell (limit~1.0 in van~Hoof's
notation) it would become $-0.125\pm0.070$.
If a constant surface brightness disc is adopted the figures would be
$-0.108\pm0.071$.

\subsection{Gaussian deconvolution versus direct measurements at the 10\% level}

\begin{figure}
  \resizebox{\hsize}{!}{\includegraphics{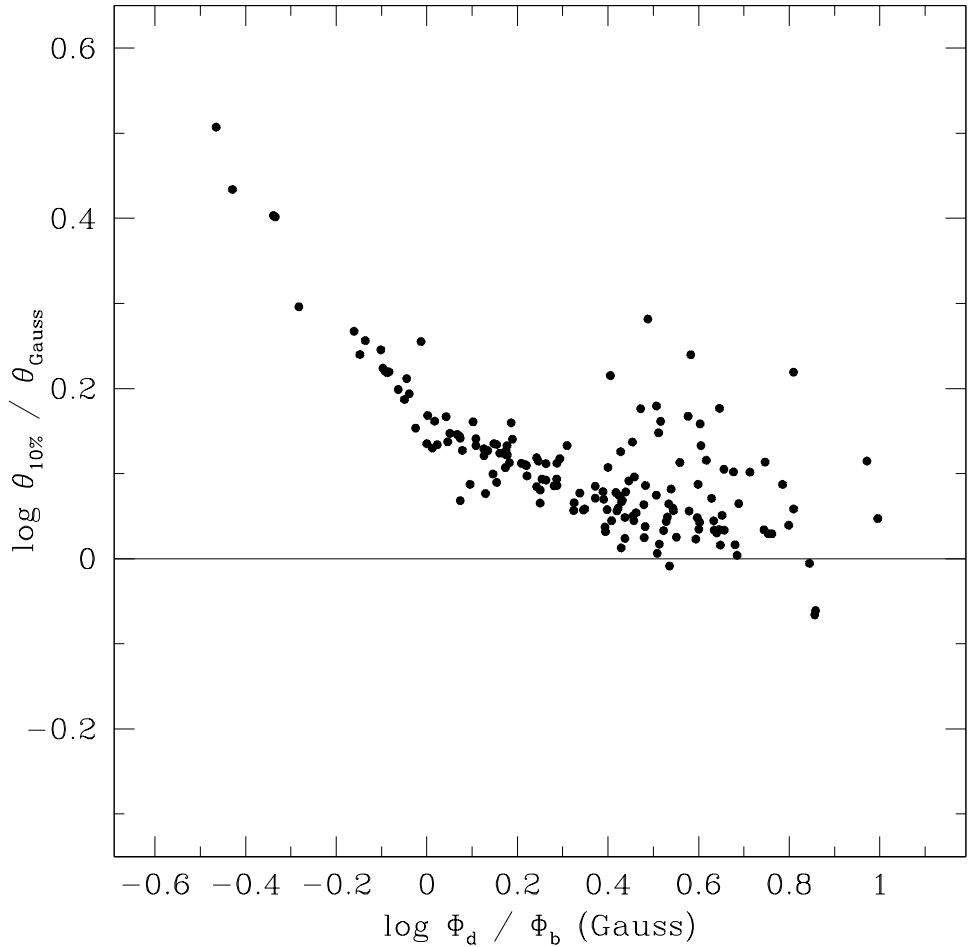}}
  \caption{Comparison of the results from the 10\% contour measurements and
           the Gaussian deconvolution. The ratio of the angular
           diameters, $\Theta$, derived from the two methods is plotted versus 
           the ratio of the deconvolved FWHM, $\Phi_{\rm d}$, to
           the resolution size, $\Phi_{\rm b}$, obtained from the Gaussian
           fit. A constant emissivity sphere model has been adopted 
           while deriving $\Theta$ in the Gaussian deconvolution.}
  \label{g_10}
\end{figure}

Fig.~\ref{g_10} compares the results from the Gaussian deconvolution with the
10\% contour measurements. Similarly as Fig.~\ref{g_sm}, it shows 
the ratio of the PN diameters versus the
Gaussian $\Phi_{\rm d}/\Phi_{\rm b}$ ratio. 
Again, the plotted values are geometric means from 
the measurements in Table~1.
While obtaining the diameters
from the Gaussian method the constant emissivity sphere model has been adopted.
Note that the diagram in Fig.~\ref{g_10} depends on the adopted surface
brightness distribution more significantly than Fig.~\ref{g_sm}. 
The general pattern of the diagram remains the same but
the points can shift vertically.

Fig.~\ref{g_10} shows an increasing discrepancy between the 10\% diameter 
and the diameter from the Gaussian deconvolution with 
the decreasing value of $\Phi_{\rm d}/\Phi_{\rm b}$.
This is an expected effect and is due to an overestimate of the PN diameter 
from the direct measurements for partially resolved PNe. 
Note, however, that a tight correlation seen in Fig.~\ref{g_10} 
for $\log \Phi_{\rm d}/\Phi_{\rm b}\,\la\,0.0$ is partly spurious.
For $\Phi$ approaching $\Phi_{\rm b}$ the y-axis in Fig.~\ref{g_10} tends 
to plot a reversal of the value plotted in the x-axis.

For small nebulae the diameters from the 10\% contour would be more accurate
if determined from deconvolved images. We have not attempted to apply this
rather cumbersome procedure to our images. Instead we have found that a
significant improvement of the 10\% diameter can be obtained using a formula
analogous to the standard FWHM Gaussian deconvolution (Eq.~(1)) 
but relevant to a 10\% level of the maximum, i.e.
\begin{equation}
\Theta_{10\%,\rm d} = \sqrt{\Theta_{10\%}^2 - (1.823\,\Phi_{\rm b})^2}.
\end{equation}
We will call $\Theta_{10\%,\rm d}$ the deconvolved 10\% diameter.

\begin{figure}
  \resizebox{\hsize}{!}{\includegraphics{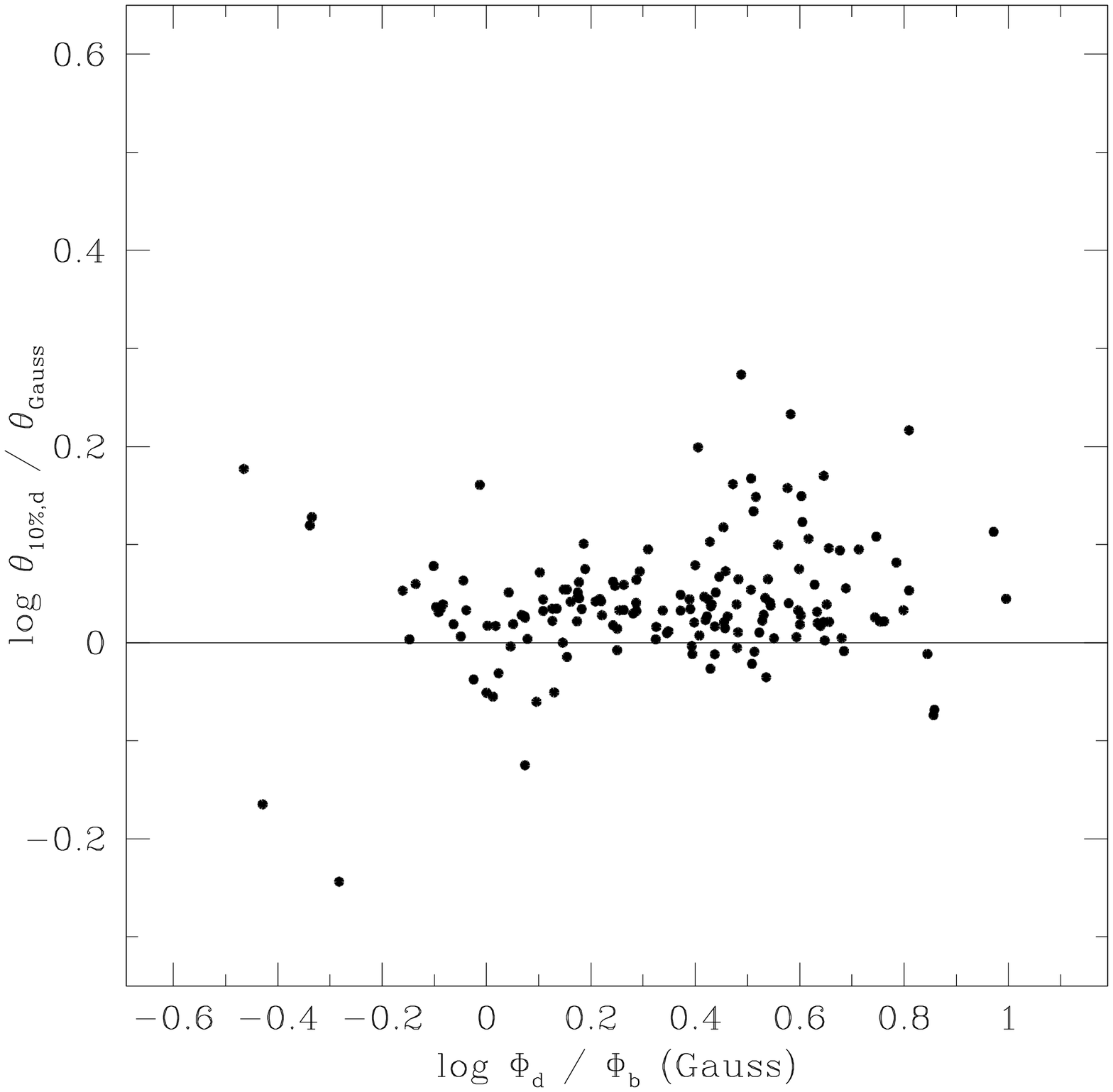}}
  \caption{The same as Fig.~\ref{g_10} but using the deconvolved 10\% 
           diameter, $\Theta_{10\%,\rm d}$, calculated from Eq.~(2).}
  \label{g_10_d}
\end{figure}

Fig.~\ref{g_10_d} shows the same as Fig.~\ref{g_10} but
using $\Theta_{10\%,\rm d}$ instead of $\Theta_{10\%}$. 
Unlike in Fig.~\ref{g_10}, the
distribution of the points in Fig.~\ref{g_10_d} does not show 
any significant trend with $\Phi_{\rm d}/\Phi_{\rm b}$.
This demonstrates that Eq.~(2) allows to significantly 
improve the reliability of the PN dimension estimate from 
the 10\% contour measurements for small, partially resolved nebulae.
$\Theta_{10\%,\rm d}$ seems to be an estimate of compact PN diameters as
reliable as that from the Gaussian deconvolution. 

However, as can be seen from Fig.~\ref{g_10_d},
the Gaussian deconvolution tends to underestimate 
the PN dimensions compared to the 10\% deconvolved diameters. 
Note that the value of this systematic difference depends on
the surface brightness distribution adopted in the Gaussian method. It is
smallest for the constant emissivity sphere, i.e. for the case presented in
Fig.~\ref{g_10_d}, and largest for the infinitely thin shell model.

Fig.~\ref{g_10_d} suggests that the two methods give the most consistent
results for rather compact but not too small nebulae, i.e. having 
$-0.2\,\la\,\log \Phi_{\rm d}/\Phi_{\rm b}\,\la\,0.3$.
The mean value of $\log \Theta_{10\%,\rm d}/\Theta_{\rm Gauss}$ for these
objects in Fig.~\ref{g_10_d} is 0.028$\pm$0.041. For an
infinitely thin shell (limit~1.0 in van~Hoof's notation) 
and a constant surface brightness disc the result would be
0.159$\pm$0.043 and 0.085$\pm$0.042, respectively.

A greater discrepancy of the results for 
$\log \Phi_{\rm d}/\Phi_{\rm b}\,>\,0.3$ can be seen from Fig.~\ref{g_10_d}.
The mean values of $\log \Theta_{10\%,\rm d}/\Theta_{\rm Gauss}$ derived
for these objects are: 0.052$\pm$0.062 (sphere), 0.195$\pm$0.062 (thin
shell), and 0.115$\pm$0.062 (disc). These nebulae show resolved structures
on the images and their surface brightness distributions are usually far
from a Gaussian. Therefore a Gaussian fitting to such objects gives
uncertain, sometimes even spurious, results.

Because of low statistics, no meaningful analysis can be done for the region 
$\log \Phi_{\rm d}/\Phi_{\rm b}\,<\,-0.2$. A large scatter of the points 
is certainly due to increasing sensitivity 
of the results to the observational uncertainties 
(quadratic subtraction of two nearly equal values).

\subsection{Second-moment deconvolution versus direct measurements at 10\%
level}

\begin{figure}
  \resizebox{\hsize}{!}{\includegraphics{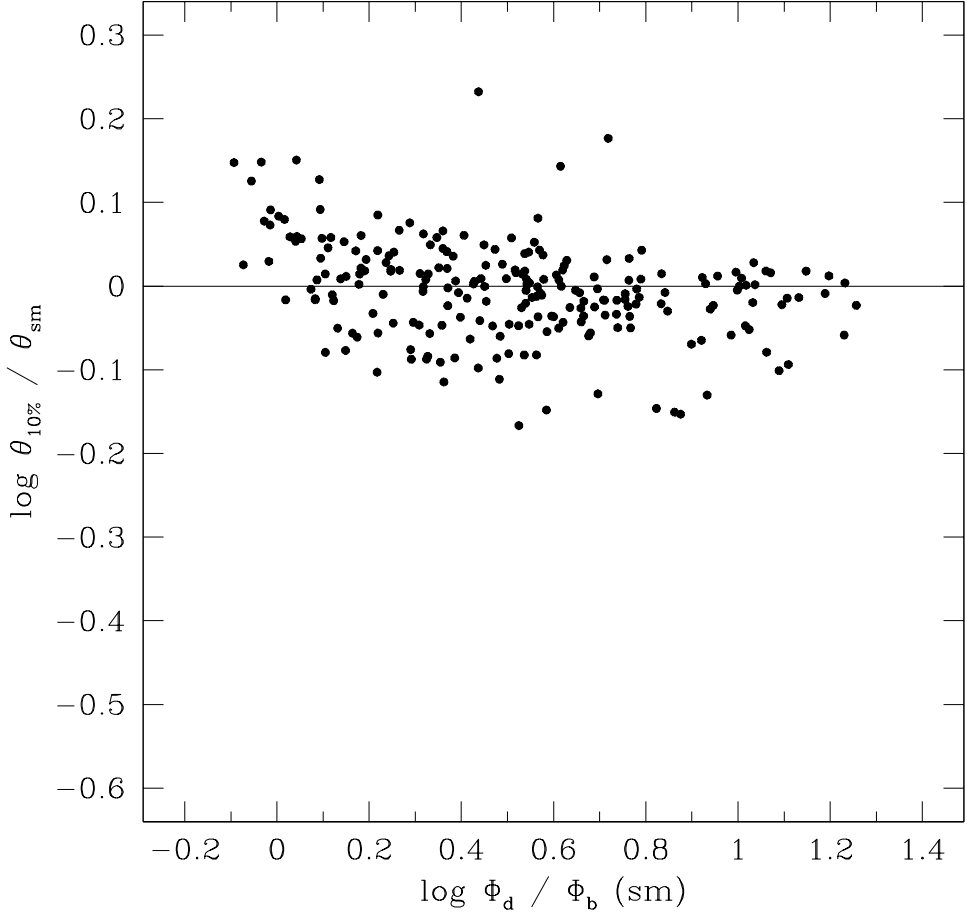}}
  \caption{Comparison of the results from the 10\% contour measurements and
           the second-moment deconvolution. The ratio of the angular
           diameters, $\Theta$, derived from the two methods is plotted versus 
           the ratio of the deconvolved FWHM, $\Phi_{\rm d}$, to
           the resolution size, $\Phi_{\rm b}$, obtained from the moment
           calculations. A constant emissivity sphere model has been adopted 
           while deriving $\Theta$ in the second-moment deconvolution.}
  \label{sm_10}
\end{figure}

\begin{figure}
  \resizebox{\hsize}{!}{\includegraphics{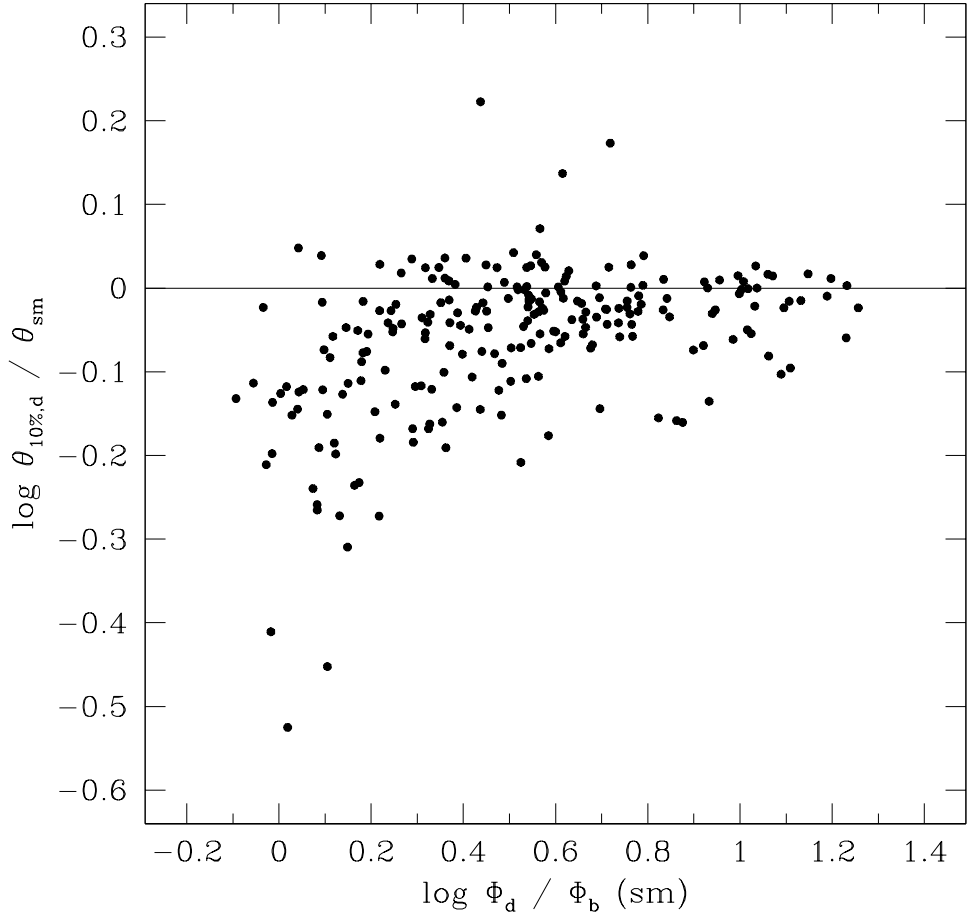}}
  \caption{The same as Fig.~\ref{sm_10} but using the deconvolved 10\% 
           diameter, $\Theta_{10\%,\rm d}$, calculated from Eq.~(2).}
  \label{sm_10_d}
\end{figure}

Figs.~\ref{sm_10} and \ref{sm_10_d} compare 
the results from the 10\% level measurements and the second-moment
deconvolution. Fig.~\ref{sm_10} is analogous to Fig.~\ref{g_10}
and uses "raw" measurements from the 10\% contour. A tendency of
overestimating diameters for compact PNe from the 10\% measurements can be
seen in Fig.~\ref{sm_10}. It is, however, much less conspicuous than 
in Fig.~\ref{g_10}. The reason is
that, as discussed in Sect.~5.1, for partially resolved PNe
the second-moment deconvolution results in 
larger diameters than the Gaussian deconvolution.

The deconvolved 10\% diameter, using Eq.~(2) and $\Phi_{\rm b}$
from the second-moment calculations, have been applied in the diagram shown
in Fig.~\ref{sm_10_d}. As the deconvolution of the 10\% diameter most
affects the results for compact nebulae the positions of these objects are
significantly shifted down in Fig.~\ref{sm_10_d} comparing to
Fig.~\ref{sm_10}.

Fig.~\ref{sm_10_d} shows that the second-moment deconvolution usually 
overestimates the PN size compared to the 10\% contour measurements. 
For rather large nebulae, i.e. having $\log \Phi_{\rm d}/\Phi_{\rm b}\,\ge\,0.3$ 
in Fig.~\ref{sm_10_d}, $\log \Theta_{10\%,\rm d}/\Theta_{\rm s-m}$ is
$-0.033\pm0.061$. The discrepancy increases for less resolved nebulae and
for $\log \Phi_{\rm d}/\Phi_{\rm b}\,<\,0.3$, 
$\log \Theta_{10\%,\rm d}/\Theta_{\rm s-m}$ is $-0.132\pm0.115$.

As the conversion factors in the second-moment deconvolution are independent
of $\Phi_{\rm d}/\Phi_{\rm b}$ adopting the infinitely thin shell model would
shift all the points in Fig.~\ref{sm_10_d} upwards by factor 0.111. 
For the uniform disc model the shift would be of factor 0.048. By the
same factors would respectively increase the mean values of
$\log \Theta_{10\%,\rm d}/\Theta_{\rm s-m}$ (obviously, the standard
deviations would remain the same).

\subsection{Testing the methods on degraded PN images}

One of the main problems in the present study is to find which method(s) gives
reliable diameter estimates for partially resolved nebulae.
The analysis discussed in the previous subsections shows that the
deconvolved 10\% diameter and the Gaussian deconvolution give relatively 
consistent results down to rather poorly resolved objects, although 
the Gaussian deconvolution 
systematically underestimates the PN diameter compared to the measurements
of the 10\% contour. The value of this underestimate depends on the model nebula
adopted in the Gaussian deconvolution. On the other hand, 
the second-moment deconvolution seems
to become unstable for compact objects as suggested by an increasing scatter
of points in Fig.~\ref{sm_10_d} for less and less resolved nebulae. However,
we cannot be sure that this interpretation is correct. Nor we can
say which method and when breaks to give reliable results. The reason is
that we can compare only the results from different methods but we do not
know what real diameters of poorly resolved nebulae are.

We have attempted to clarify the above problems by applying the methods of
diameter measurements to some of our images artificially degraded to worse and
worse seeing conditions, i.e. convolved with a larger and larger
Gaussian profile. From the original images we know the true PN sizes so we
can compare them to the results from different methods applied to
less and less resolved PN images.

Although the idea is simple it could have been applied to a
limited number of our images only. Obviously we have to consider rather well
resolved nebulae but not too much. 
The object has to be several times smaller than the image size. 
Otherwise the image frame can impose an important cutoff when the image is
convolved with a Gaussian having FWHM comparable to the PN size. The nebula
has to be well exposed so its image keeps a good contrast while beeing
spread over larger and larger surface. There cannot be too many bright field
stars, otherwise they merge with the PN image in the convolution process.
It is also desirable that among the objects investigated different morphologies
are represented.

We present the results of the procedure applied to images of 5 PN, i.e. (in paranthesis
usual names and morphological descriptions are given) 021.8--00.4 (M\,3-28; 
bipolar, butterfly-like), 174.2--14.6 (H\,3-29; ring-like but strongly
asymmetric with one strong blob), 285.7--14.9 (IC\,2448; round with a halo),
286.3--04.8 (NGC\,3211; ring-like with two blobs), and 345.2--08.8 (Tc\,1; round
with an irregular halo). The selected objects
have diameters between 10 and 20 arcsec. The original
images have been obtained with a seeing varying from 0.5 to 2.5 arcsec. 
We have used our deconvolved 10\% diameters  from original
images as true PN diameters for comparison with the results from degraded
images (note that the deconvolution has
introduced only minor corrections to the 10\% contour measurements in these
cases). The original images have been convolved with a Gaussian having FWHM
increasing from 2.5 to 20.0 arcsec. Then our three methods have been
applied to the resulting images.

We note that the results from measurements on the degraded images reproduce
very well the general trends seen in Figs.~\ref{g_sm}--\ref{sm_10_d}.

\begin{figure}
  \resizebox{\hsize}{!}{\includegraphics{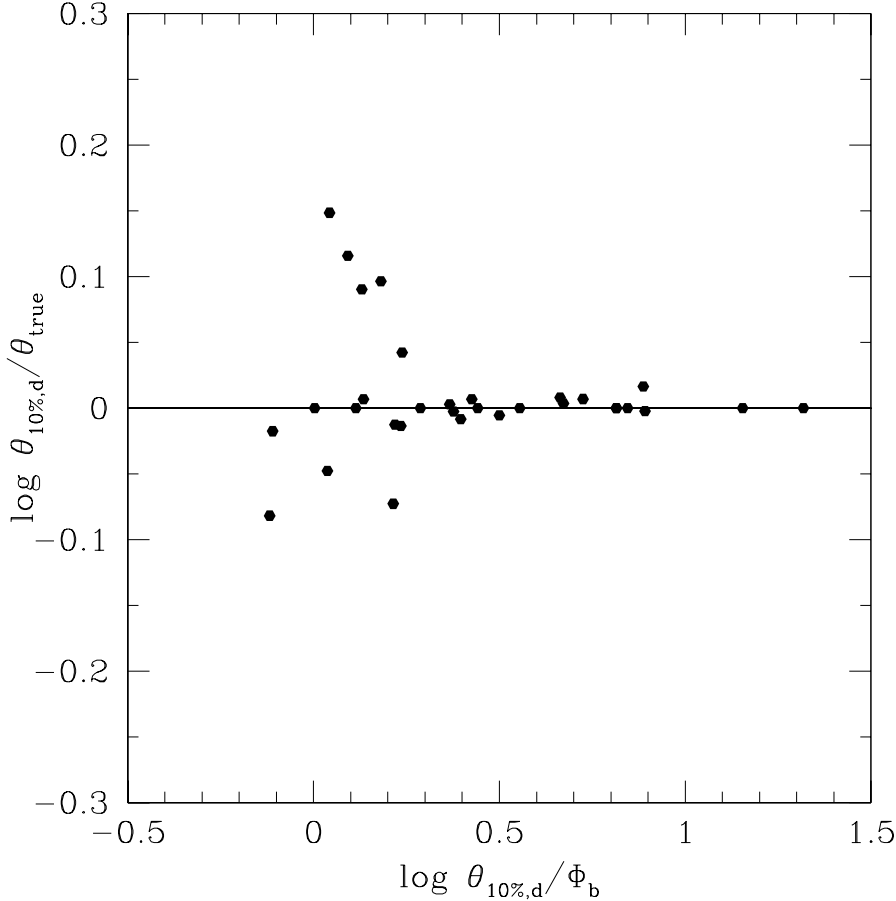}}
  \caption{Results of the 10\% contour measurements on the degraded PN
           images. $\Theta_{10\%,\rm d}$ and $\Phi_{\rm b}$ are the deconvolved 10\% 
           diameter and the seeing size, respectively, measured from the
           degraded images. $\Theta_{\rm true}$ is the deconvolved 10\%
           diameter from the original (non-degraded) images.}
  \label{10d_vs_true}
\end{figure}

In Fig.~\ref{10d_vs_true} the results of the deconvolved 10\% diameter
method applied to the degraded images are compared to the deconvolved 10\%
diameter from the original (non-degraded) images ($\Theta_{\rm true}$). From
this figure we can conlude that the deconvolved 10\% diameter
measures very well the PN size down to 
$\log \Theta_{10\%,\rm d}/\Phi_{\rm b}\,\simeq\,0.3$. This limit is
practically equivalent to $\log \Phi_{\rm d}({\rm Gauss})/\Phi_{\rm b}\,\simeq\,0.05$.
Below this value the method becomes unstable.

\begin{figure}
  \resizebox{\hsize}{!}{\includegraphics{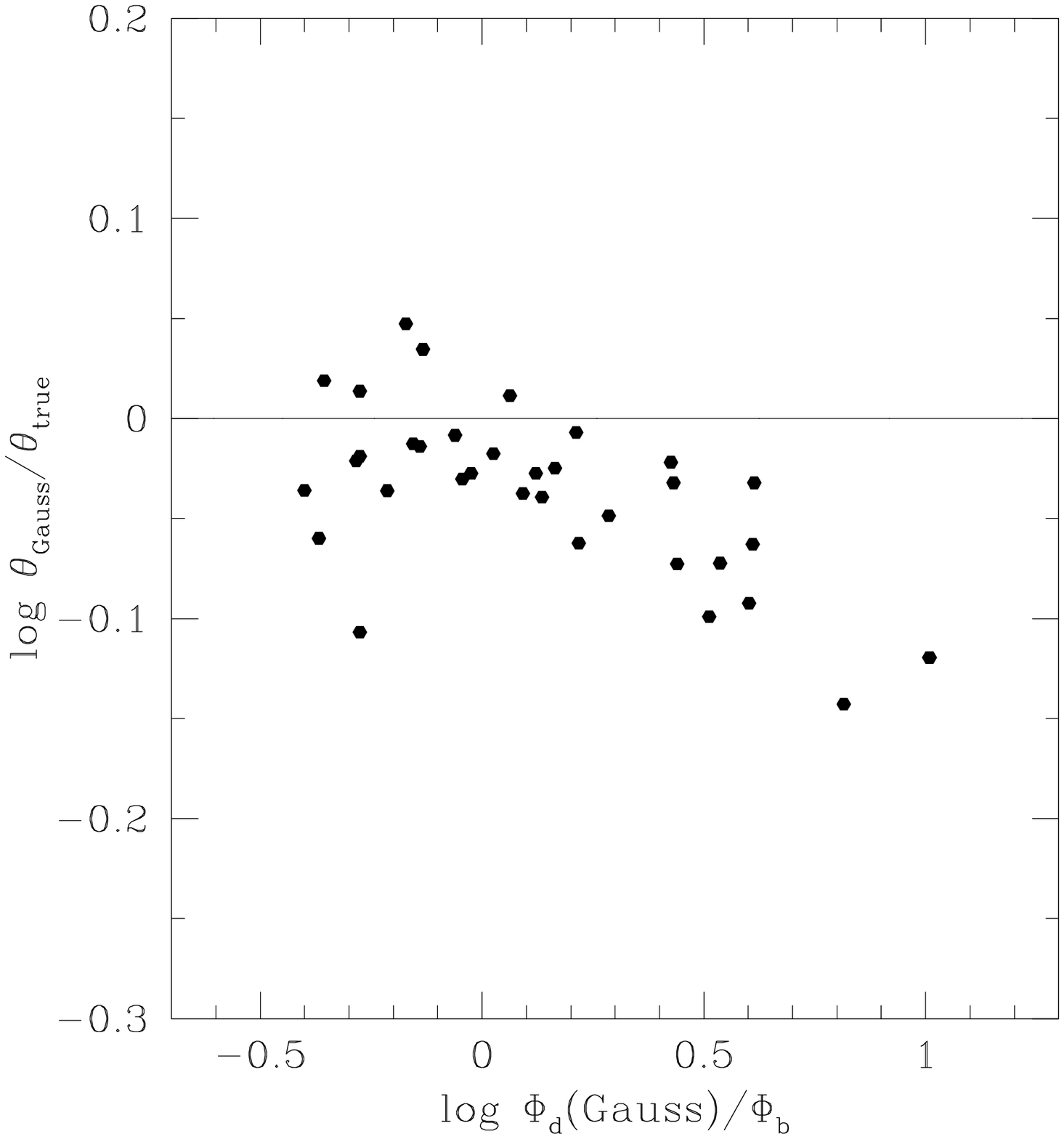}}
  \caption{The same as Fig.~\ref{10d_vs_true} but for the Gaussian
           deconvolution adopting a constant emissivity sphere model.
           $\Theta_{\rm true}$ is the deconvolved 10\%
           diameter from the original (non-degraded) images.}
  \label{gauss_vs_true}
\end{figure}

Results of the Gaussian deconvolution applied to the degraded images are
presented in Fig.~\ref{gauss_vs_true}. A constant emissivity sphere model
has been adopted. We see that for well resolved nebulae the Gaussian
deconvolution underestimates the PN diameter. This is consistent with the
trend seen in Fig.~\ref{g_10_d}. For $\log \Phi_{\rm d}/\Phi_{\rm b}\,\la\,0.3$
the underestimate is smaller, again similar as in Fig.~\ref{g_10_d}. 
For all the points with 
$\log \Phi_{\rm d}/\Phi_{\rm b}\,<\,0.3$ in Fig.~\ref{gauss_vs_true} the
mean value and the standard deviation of 
$\log \Theta_{\rm Gauss}/\Theta_{\rm true}$
are $-0.021\pm0.032$. This result is very close to the value
$-0.028\pm0.041$
of $\log \Theta_{\rm Gauss}/\Theta_{10\%,\rm d}$
found in Sect. 5.2 for objects having
$-0.2\,\la\,\log \Phi_{\rm d}/\Phi_{\rm b}\,\la\,0.3$
in Fig.~\ref{g_10_d}.
For $\log \Phi_{\rm d}/\Phi_{\rm b}\,\la\,-0.2$ in Fig.~\ref{gauss_vs_true}
the scatter increases which
suggests that the Gaussian deconvolution becomes then unstable. Comparison
of Figs.~\ref{gauss_vs_true} and \ref{10d_vs_true} seems to indicate that
the Gaussian deconvolution gives somewhat more stable results than the
deconvolved 10\% diameter for objects having
$\log \Theta_{10\%,\rm d}/\Phi_{\rm b}\,\le\,0.3$ or,
equivalently, $\log \Phi_{\rm d}({\rm Gauss})/\Phi_{\rm b}\,\le\,0.05$. A better
statistics would be required to clarify this point.

\begin{figure}
  \resizebox{\hsize}{!}{\includegraphics{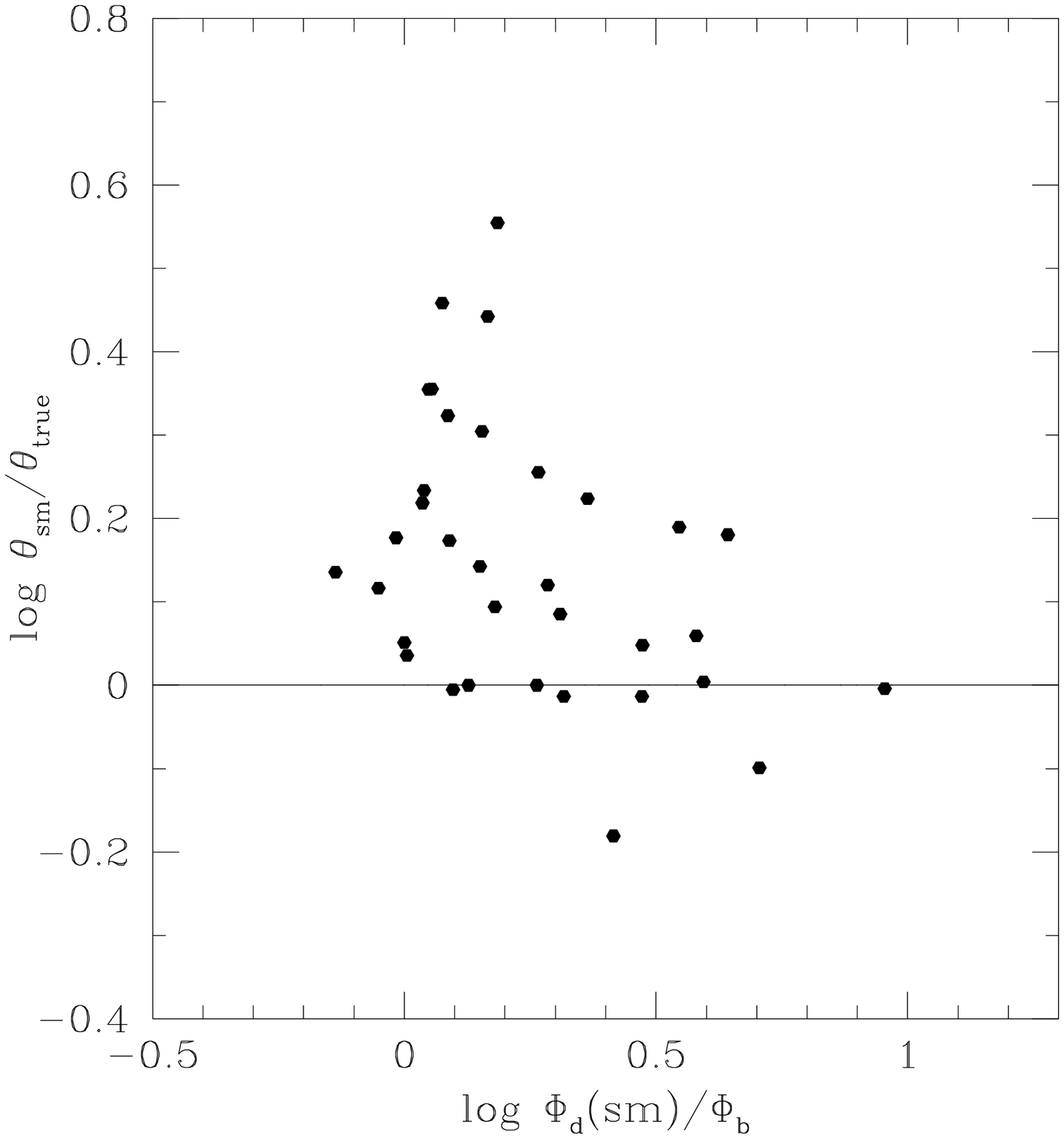}}
  \caption{The same as Fig.~\ref{10d_vs_true} but for the second-moment
           deconvolution adopting a constant emissivity sphere model.
           $\Theta_{\rm true}$ is the deconvolved 10\%
           diameter from the original (non-degraded) images.}
  \label{sm_vs_true}
\end{figure}

The results of the test of the second-moment deconvolution are presented in
Fig.~\ref{sm_vs_true}. A constant emissivity sphere model
has been adopted. The figure shows that the second-moment deconvolution
gives uncertain estimates of the PN diameters, especially for partially
resolved nebulae ($\log \Phi_{\rm d}/\Phi_{\rm b}\,<\,0.3$). Adopting a
shell or disc model in the deconvolution would decrease the systematic
difference between $\Theta_{\rm s-m}$ and $\Theta_{\rm true}$ in Fig.~\ref{sm_vs_true}
but the uncertainity (scatter of the points) would remain the same. Our
experience suggests that this method is sensitive to faint outer structures
(halo, outer butterfly-like filaments) particularly in partially resolved
nebulae.

\section{PN diameters}

One of the primary goals of our work is to obtain a homogeneous set of 
PN dimensions, a set of results which have, as far as possible, the same
physical meaning for all the objects in our sample. As a primary method for
deriving the PN dimensions we have adopted the direct measurements at the
10\% level of the peak surface brightness. This seems to be the only reasonable
method for well resolved nebulae. In order to account for the finite
resolution in the case of smaller objects we have used the deconvolved 10\%
diameters, $\Theta_{10\%,\rm d}$, instead of $\Theta_{10\%}$. As discussed in
Sect.~5.4 $\Theta_{10\%,\rm d}$ is reliable down to 
$\log \Theta_{10\%,\rm d}/\Phi_{\rm b}\,\simeq\,0.3$.
For the smallest objects when results from any method become more and
more uncertain we have used $\Theta_{10\%,\rm d}$ and the results from 
the Gaussian deconvolution.
Following discussion in Sect.~5.4 we have prefered not to use 
the results from the second-moment deconvolution.
In more detail our prescription is as follows.

For all the objects with $\log \Theta_{10\%,\rm d}/\Phi_{\rm b}\,\ge\,0.3$
we have adopted $\Theta_{10\%,\rm d}$. 
We have used $\Phi_{\rm b}$ from Gaussian fit.
For more compact objects, i.e.
when $\log \Theta_{10\%,\rm d}/\Phi_{\rm b}\,<\,0.3$
we have adopted a mean
value from $\Theta_{10\%,\rm d}$ and $\Theta$ resulting from the Gaussian
deconvolution adopting the constant emissivity sphere model. In order to
correct for the systematic underestimate of the Gaussian method, as discussed
in Sects.~5.2 and 5.4, the Gaussian $\Theta$ have been multiplied by factor
1.06 before deriving the mean value.

The results are presented in Table~2. Columns (1) and (2) give the PNG
numbers and the usual names, respectively. The diameters derived according
to the above prescription are given in column (3). In the case of partially
resolved objects having $\log \Theta_{10\%,\rm d}/\Phi_{\rm b}\,<\,0.3$
only one value is given which is the
geometric mean from the two dimension measurements.
A colon indicates compact 
objects for which the Gaussian deconvolution resulted in a diameter
different by more than 30\% from the deconvolved 10\% diameter. Objects
practically unresolved in our images are marked with "stellar". For a few
objects having two independent measurements in Table~1 mean values
derived from them are given in Table~2.


\begin{figure}
  \resizebox{\hsize}{!}{\includegraphics{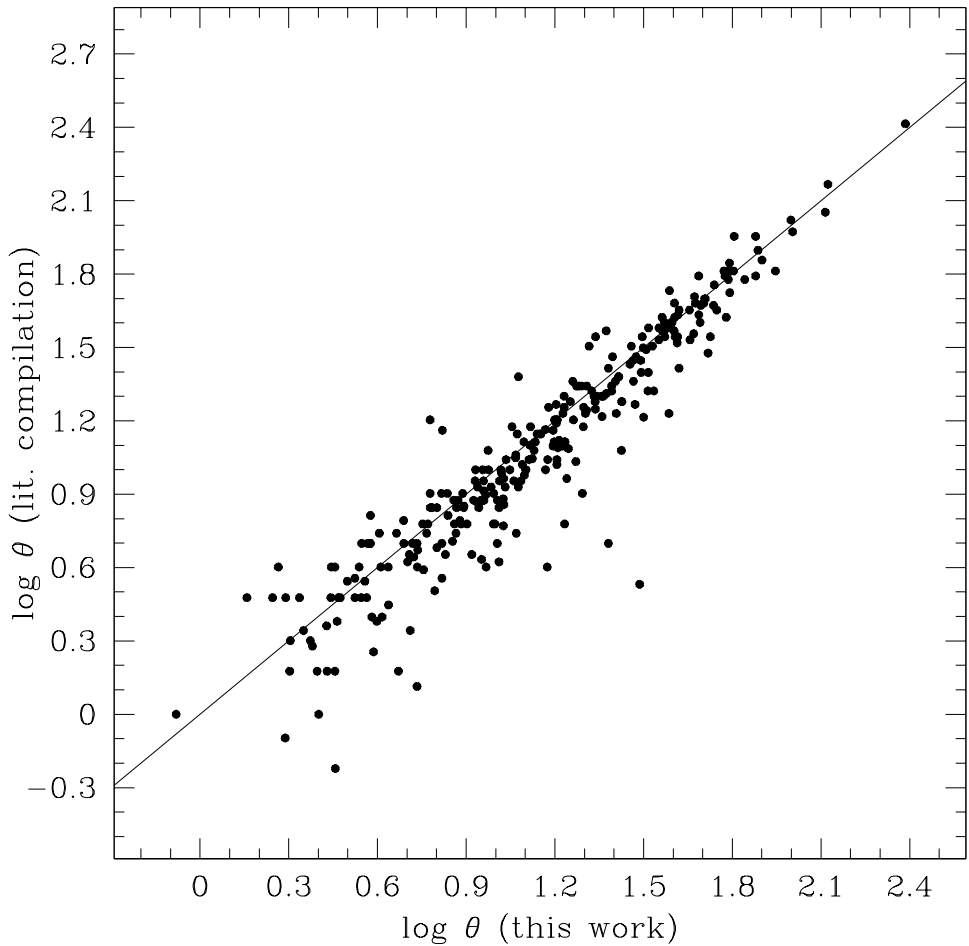}}
  \caption{Comparison of the PN diameters from the literature compilation
           (see text)
           with the diameters from this work (geometric means from the two
           dimensions given in Table~2). Solid line: 1 to 1 relation.}
  \label{lit_tw}
\end{figure}

We have compared the results from Table~2 with the diameters from our data
base resulting from a compilation of the data published in the literature.
This data base has originally been made for purposes of the Strasbourg-ESO
catalogue (Acker et~al. \cite{gpnc}). Its updated version
has been described and used in Si\'odmiak \& Tylenda
(\cite{st}). The results of the comparison are shown in Fig.~\ref{lit_tw}. In most
cases the agreement is quite good although our measurements tend to give
somewhat larger diameters than the values adopted from the literature. 
The mean value of the ratio of our diameters (from Table~2) to the compiled diameters 
is 1.24 with a standard deviation of 0.71. In some
cases, however, the discrepancy is significant. For small nebulae a
large scatter of the points in Fig.~\ref{lit_tw} is not surprising. Any measurement,
VLA or optical, of a partially resolved nebulae gives an uncertain result.
Significant discrepancies in the case of larger nebulae usually concern
bipolar nebulae having a compact core, e.g. IC\,4634 (000.3+12.2),
M\,3-28 (021.8--00.4), M\,2-48 (062.4--00.2), He\,2-104 (315.4+09.4), 
Hb\,5 (359.3--00.9). The published diameters from VLA or optical measurements
apparently referred to the core whereas the 10\% contour from our images
encompassed a significant portion of the outer bipolar structures.

\begin{figure}
  \resizebox{\hsize}{!}{\includegraphics{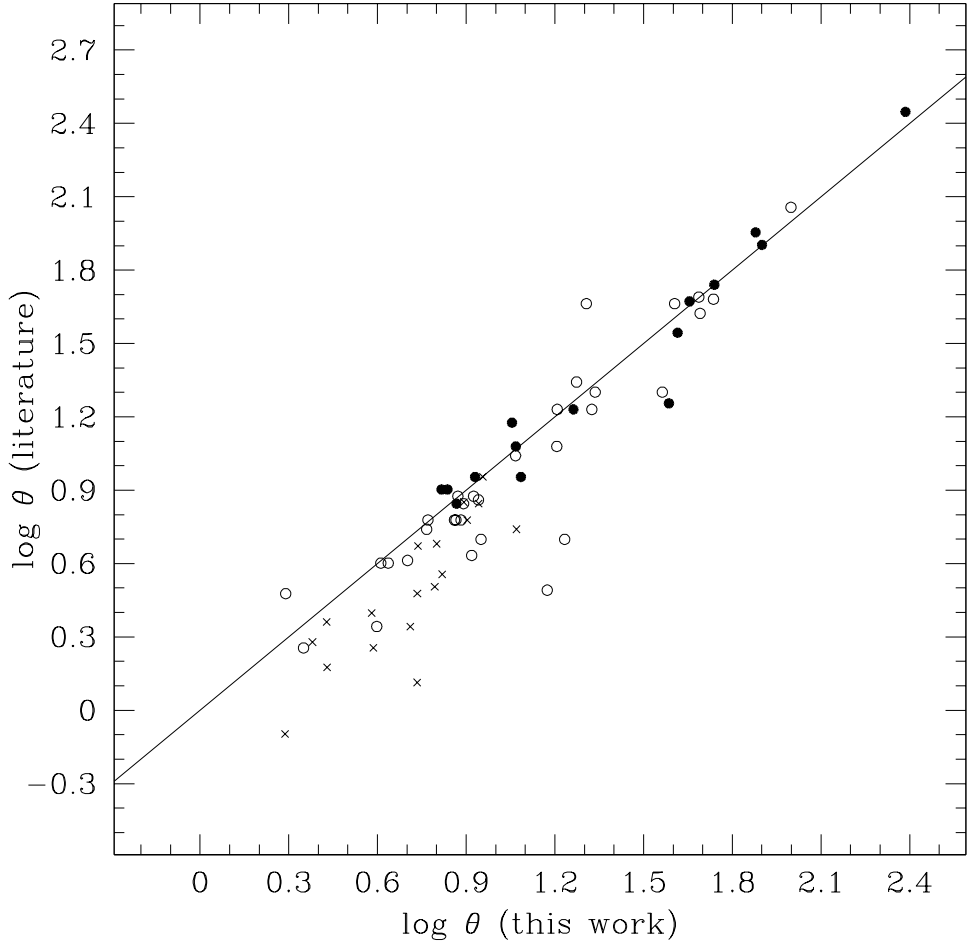}}
  \caption{Comparison of the PN diameters from Aaquist \& Kwok (\cite{ak})
           (crosses) and Zijlstra et~al. (\cite{zpb}) 
           (full circles -- from 10\% contour, 
            open circles -- from Gaussian deconvolution)
           with the diameters from this work (geometric means from the two
           dimensions given in Table~2). Solid line: 1 to 1 relation.}
  \label{ak_zpb_tw}
\end{figure}

In Fig.~\ref{ak_zpb_tw} we have compared the diameters from 
Table~2 with the results of VLA surveys of 
Aaquist \& Kwok (\cite{ak}) and Zijlstra et~al. (\cite{zpb}).
Aaquist \& Kwok observed mainly compact PNe. They have applied three methods
for deriving the PN diameters but it is not clear which method(s) has been
used to measure a particular object. As can be seen from Fig.~\ref{ak_zpb_tw}, our
diameters are systematically larger than those of Aaquist \& Kwok, on
average by factor 1.7. The agreement of our results with those of Zijlstra
et~al. is better especially if nebulae greater than $\sim$10 arcsec are
considered and the diameter of Zijlstra et~al. resulted from their 10\%
contour (full circles in the figure). 
In four cases the discrepancy is larger than factor 2. All the
cases concern bipolar nebulae. For three of them, i.e already mentioned
M\,3-28, M\,2-48, as well as NGC\,2440 (234.8+02.4), our diameters are
larger and the reason has been explained above. In the case of M\,2-9 our
diameter is significantly smaller and it seems that Zijlstra et~al. measured
the angular extend of the elongated, bipolar lobes whereas our value is a
geometric mean from the major and minor axis measurements.

\section{Discussion and conclusions}

We have measured angular dimensions for 312 PNe. As a primary definition of
the PN size we have adopted the extension of nebular emission up to 
the 10\% level of the maximum surface brightness. This seems to be a
reasonable choice for extended, well resolved nebulae.
For partly resolved objects deconvolution methods have to be used.
So far the Gaussian deconvolution has primarily been applied.

We have, however, found that for compact nebulae the 10\% contour measurements 
can also give satisfactory results provided that they are
corrected for the finite resolution using Eq.~(2).
As has been shown in Sect.~5.2 and 5.4 this simple deconvolution of the 10\%
contour measurments makes the resulting diameters reliable down to nebulae
as compact as $\log \Theta_{10\%,\rm d}/\Phi_{\rm b}\,\simeq\,0.3$ (or
equivalently $\log \Phi_{\rm d}({\rm Gauss})/\Phi_{\rm b}\,\simeq\,0.05$).
The derivation of $\Theta_{10\%,\rm d}$ has several
advantages over the deconvolution methods. It is simpler to apply. 
No assumption has to be made about the surface brightness distribution.
It is a more universal method as it can be applied to compact and 
partially resolved PNe as well as to
extended and well resolved nebulae where the deconvolution methods 
usually fail in practice.

The Gaussian deconvolution gives fairly reliable and stable results for
partly resolved nebulae. However it systematically underestimates the
diameters compared to the 10\% contour measurements. The difference depends
on the model nebula adopted in the Gaussian
deconvolution. It is only 5-7\% for the constant emissivity sphere but
increases up to 45\% for a thin shell.
This result means that in catalogues and data bases the dimensions of
compact PNe, usually derived from the Gaussian deconvolution, 
are probably underestimated compared to those of extended nebulae. It is
difficult to estimate this systematic difference.
The deconvolution procedures used in observational
surveys, like in Aaquist \& Kwok (\cite{ak}) or Zijlstra et~al. (\cite{zpb}),
were significantly simplified compared to that of van~Hoof (\cite{vh}).
Usually the same value of the conversion factor has been used for all the objects. 
This certainly introduced additional uncertainties in the final results.

The second-moment deconvolution appeared to give results often uncertain and 
in poor agreement with the results from the other two methods, 
in particular for compact nebulae. 

Our study does not solve definitevely the problem of measuring the PN dimensions.
It makes however an important step in the field in the sense that it gives a
quantity which means physically the same and is comparable for all the objects 
in our sample. One has, however, to remember that the (faint) nebular emission can
extend well beyond the diameters given in Table~2. This can be seen for many
objects in our sample by comparing column (9) and column (8) in Table~1.

\begin{acknowledgements}
 We are very grateful to the referee (J.J. Condon) whose comments resulted
 in a significant improvement of the work reported in this paper.
 This work has partly been supported from a grant No.
 2.P03D.020.17 financed by the Polish State Committee for Scientific Research.
\end{acknowledgements}

\end{document}